\documentclass[showpacs,showkeys,aps,preprint]{revtex4}
\usepackage{graphicx}
\usepackage{dcolumn}
\usepackage{bm}
\begin{document}
\newcommand{\ti}[1]{\mbox{\tiny{#1}}}
\newcommand{\im}{\mathop{\mathrm{Im}}}
\def\be{\begin{equation}}
\def\ee{\end{equation}}
\def\bea{\begin{eqnarray}}
\def\eea{\end{eqnarray}}

\title{Generating static perfect-fluid solutions of Einstein's equations}

\author{Hernando Quevedo$^{1,2}$  and Saken Toktarbay$^3$}
\email{quevedo@nucleares.unam.mx,saken.yan@yandex.com}
\affiliation{
$^1$Instituto de Ciencias Nucleares, Universidad Nacional Aut\'onoma de M\'exico, AP 70543, M\'exico, DF 04510, Mexico\\
$^2$Dipartimento di Fisica and ICRA, Universit\`a di Roma ``La Sapienza", I-00185 Roma, Italy\\
$^3$Department of Theoretical and Nuclear Physics, Al-Farabi Kazakh National University,  Almaty 050040, Kazakhstan
}

\date{\today}

\begin{abstract}
We present a method for generating exact interior solutions of Einstein's equations in the case of static and axially symmetric perfect-fluid spacetimes. 
The method is based upon a transformation that involves the metric functions as well as the density and pressure of the seed solution. In the limiting vacuum case, 
it reduces to the Zipoy-Voorhees transformation that can be used to generate metrics with multipole moments. All the metric functions of the new solution can be calculated explicitly from the seed solution in a simple manner. The physical properties of the resulting new solutions are shown to be completely different from those of the seed solution.

\end{abstract} 

\maketitle

\section{Introduction}
\label{sec:int}

General relativity is a theory of gravity and as such should be able to describe the field of all possible physical configurations in which gravity is involved. 
All the information about the gravitational field should be encoded in the metric tensor which must be an exact solution of Einstein's equations. In this work, we will focus on the study of the gravitational field generated by compact objects like planets or stars, which can be considered as independent of time. In this case,  the problem of describing the  complete field generated by the source  can be split into two correlated problems, namely, the interior and the exterior rotating fields. To handle the corresponding field equations, one usually assumes axial symmetry with respect to the rotation axis. 
It is well known that the exterior field of an arbitrarily rotating mass can be described by the Kerr spacetime \cite{kerr63}.
As for the interior field, the situation is more complicated. In fact, a major actual problem of classical general
relativity consists in finding a physically reasonable interior solution for the exterior Kerr metric, which could be applied to describe the interior field of realistic compact objects. 
One usually assumes that the matter inside the object can be described by a rotating perfect fluid. Since the discovery of the Kerr metric in 1963, many attempts have been made to find the corresponding  exact perfect-fluid interior solution. It seems that the rotation parameter for a perfect fluid leads to  important difficulties in the context of Einstein's equations. In this work, we will concentrate on the problem of perfect-fluid solutions, without considering the rotation parameter. We expect that the understanding of the symmetry properties of prefect-fluid solutions will help to incorporate later the rotation parameter. 

Although there exists quite a large number of exact solutions \cite{solutions}, only a few can be considered as physically meaningful. This type of solutions are usually classified in terms of their symmetry. Consider, for instance, spherically symmetric spacetimes. According to Birkhoff's theorem, the exterior field is uniquely determined by the Schwarzschild solution. For the interior field, however, a quite large number of exact solutions are known. In fact, the explicit form of the interior solutions depends on the model used to determine the energy-momentum tensor of the source. In addition, if we limit ourselves to the case of perfect-fluid sources, the search for exact solutions requires the knowledge of an additional equation which is usually taken as the equation of state of the fluid. The explicit form of the interior solution depends heavily on the properties of the equation of state. Recently, several generating techniques have been proposed that allow one to obtain all spherically symmetric perfect-fluid interior solutions with and without equations of state \cite{lake03,ramvis02,hop08,sem11}. 

In the case of stationary axisymmetric spacetimes, only a few solutions are known for the interior of a rigidly rotating perfect fluid. In  
\cite{wahl68,vai77,kram84} several exact solutions were found which, however, are characterized by negative pressures. In \cite{seno87,seno96} more realistic interior solutions were obtained with physically plausible equations of state for rigidly rotating sources. Nevertheless, the problem of finding the solution for the corresponding exterior field remains open because the attempts to solve directly the resulting vacuum field equations, together with the energy conditions and the matching conditions, have not been very successful.  In view of this situation, it seems reasonable to explore other methods. In particular, we believe that 
the methods for generating new solutions from known ones could give some new insight into the problematic \cite{solutions}. 
The present work can be considered as a first step in this direction.

In this work, we focus on non-rotating perfect-fluid configurations. We propose a line element which is specially adapted to the investigation of the symmetry properties of the corresponding field equations. In fact, we will show that a particular symmetry corresponds to a transformation that can be used to generate new exact solutions. This paper is organized as follows. In Sec. \ref{sec:lel}, we propose  for static axisymmetric interior spacetimes a particular metric whose field equations can be written in a relatively compact form that resembles the vacuum case. In Sec. \ref{sec:tra}, we prove a theorem about a transformation that can be used to generate new exact solutions, and can interpreted as a generalization of the Zipoy-Voorhees transformation for the vacuum limiting case. In Sec. \ref{sec:ex}, we present some explicit examples of the procedure proposed here to generate new perfect-fluid solutions of Einstein's equations. Finally, in Sec. \ref{sec:con}, we discuss our results and comment some open problems.

\section{Line element and field equations}
\label{sec:lel}

We will consider a static axially symmetric spacetime with Killing vectors $\xi_t = \partial_t$ and $\xi_\varphi=\partial_\varphi$, where $t$ is the time-coordinate and
$\varphi$ is the azimuthal angle. This implies that $\partial g_{\alpha\beta}/\partial t = \partial g_{\alpha\beta}/\partial \varphi=0$. For the remaining spatial coordinates we choose $\theta$ as the polar angle and $r$ as a radial-like coordinate. As for the line element for this type of spacetimes, several choices are possible \cite{solutions}. In the case of interior solutions, a particular choice \cite{ster03} has been intensively used to derive and analyze approximate numerical solutions.
In principle, for a given set of symmetry conditions and coordinates, all possible line elements should be equivalent. However, the point is that the structure of the field equations depends on the explicit form of the line element and, therefore, a particular choice might be more suitable for the investigation of the field equations structure. In this work, we propose the following line element 
\be
ds^2 = e^{2\psi} dt^2 - e^{-2\psi}\bigg[e^{2\gamma}\left(\frac{dr^2}{h} + d\theta^2\right) + \mu^2 d\varphi^2\bigg]\ ,
\label{lel}
\ee
to study the structure of the field equations with a perfect-fluid source. Here $\psi=\psi(r,\theta)$, $\gamma=\gamma(r,\theta)$, $\mu=\mu(r,\theta)$, and $h=h(r)$. Moreover, we use geometric units with $G=c=1$. The Einstein equations for a perfect fluid with 4-velocity $U_\alpha$, density $\rho$, and pressure $p$  
\be
R_{\alpha\beta} - \frac{1}{2} R g_{\alpha\beta} = 8 \pi \left[(\rho+p)U_\alpha U_\beta - p g_{\alpha\beta}\right] 
\label{eins}
\ee
for the line element (\ref{lel}) lead to a system of four independent differential equations. The main equations, relating the metric functions $\mu$, $\psi$, and $h$, can be written as 
\be
\frac{\mu_{,rr}}{\mu} +  \frac{\mu_{,\theta\theta}}{h\mu} +\frac{h_{,r}\mu_{,r}}{2h\mu} = \frac{16\pi}{h} \bar p \ ,
\label{mu}
\ee 
\be
\psi_{,rr} + \frac{\psi_{,\theta\theta}}{h} +\left(\frac{h_{,r}}{2h}+\frac{\mu_{,r}}{\mu}\right) \psi_{,r} + \frac{\mu_{,\theta} \psi_{,\theta}}{h\mu} 
= \frac{4\pi}{h}(3 \bar p + \bar\rho)\ ,
\label{psi}
\ee
where the comma represents partial differentiation with respect to the corresponding coordinates, and we have introduced the ``normalized" density $ \bar \rho$ and pressure $\bar p$ by means of
\be 
\bar \rho = \rho e^{2\gamma - 2\psi}\ ,\quad \bar p = p e^{2\gamma - 2\psi}\ .
\ee
Notice that there is no independent equation for the metric function $h(r)$. The reason is that in the line element (\ref{lel}) it is possible to introduce a new radial-like coordinate, say $dR= dr/\sqrt{h}$, which absorbs the function $h$, implying that it does not represent a genuine degree of freedom. Nevertheless, we keep  $h$ as an auxiliary function which can be chosen arbitrarily, in principle. Accordingly, if we attempt to solve the above system it is necessary to specify $h$, by using some additional criteria. This will turn to be useful for the purpose of the present work, as we will show below. As for the metric function $\gamma$, it is determined by two first order differential equations    
\be
\gamma_{,r} = \frac{1}{h\mu_{,r}^2+\mu_{,\theta}^2}\bigg\{\mu\left[\mu_{,r}\left(h\psi_{,r}^2-\psi_{,\theta}^2\right) +2\mu_{,\theta}\psi_{,\theta}\psi_{,r} +  8\pi\mu_{,r}\bar p\right]
+\mu_{,\theta}\mu_{,r\theta}-\mu_{,r}\mu_{,\theta\theta}  \bigg\}\ ,
\label{gamr}
\ee
\be
\gamma_{,\theta} = \frac{1}{h\mu_{r}^2+\mu_{,\theta}^2}\bigg\{ \mu\left[\mu_{,\theta}\left(\psi_{,\theta}^2-h\psi_{,r}^2\right) +2 h \mu_{,r}\psi_{,\theta}\psi_{,r}  - 8\pi\mu_{,\theta}\bar p \right] 
+ h\mu_{,r}\mu_{r\theta} +\mu_{,\theta}\mu_{,\theta\theta}\bigg\}\ ,
\label{gamt}
\ee
which can be integrated by quadratures once the main field equations (\ref{mu}) and (\ref{psi}) are solved, and the normalized pressure $\bar p$ is known. It is important to notice that if we introduce the differential equations (\ref{mu})-(\ref{gamt}) into the original Einstein equations (\ref{eins}), a second order differential equation for $\gamma$ emerges 
\be
\gamma_{,rr}+\frac{\gamma_{,\theta\theta}}{h}+\psi_{,r}^2 + \frac{\psi_{,\theta}^2}{h} + \frac{h_{,r}\gamma_{,r}}{2h}= \frac{8\pi}{h}\bar p \ ,
\label{gam2}
\ee
which must also be satisfied. Nevertheless, it can be shown that this equation is identically satisfied if the two first order differential equations (\ref{gamr}) and 
(\ref{gamt}) for $\gamma$ and the conservation equation for the non-normalized parameters of the perfect fluid
\be 
p_{,r} = -(\rho+p)\psi_{,r}  \ ,\quad p_{,\theta} = -(\rho+p)\psi_{,\theta}
\label{claw}
\ee
are satisfied. 

Notice that the structure of the differential equations as presented above resembles the structure of the vacuum field equations. In fact, for vanishing density and pressure with $h=1$, the equation for $\mu$ is trivially satisfied, if $\mu$ is used as a radial-like coordinate ($\mu\sim r$), and the only remaining field equation is 
(\ref{psi})  for the function $\psi$ whose solution determines uniquely the remaining function $\gamma$. Since the conservation law is identically satisfied in this case, the compatibility condition following from Eqs.(\ref{gamr})-(\ref{gam2}) coincides with the main field equation (\ref{psi}). 

Notice also that the present choice of the line element leads to a particularly simple generalization of the Tolman-Oppenheimer-Volkov (TOV) equation for spherically symmetric perfect-fluid sources. This has been also obtained previously with different choices of line elements \cite{ster03}. In fact, the TOV equation coincides with the first equation given in (\ref{claw}), with $r$ being the radial spherical coordinate. The generalization to include the axisymmetric case implies a second TOV equation for the angle coordinate $\theta$.  This simple form of the TOV equations allows us to easily calculate the pressure, if $\rho$ and $\psi$ are given.

\section{The transformation}
\label{sec:tra}

The search for physically reasonable exact solutions of the above system of differential equations is in general a very complicated task. In principle, we have a system of four independent equations (\ref{mu})-(\ref{gamt}) for the six unknowns $\mu$, $\psi$, $\gamma$, $h$, $p$, and $\rho$. To close the system it is necessary to specify $h=h(r)$ and, for instance, an equation of state of the form $p=p(\rho)$. We investigated the simple case of a polynomial function $h(r)$ and a barotropic equation of state 
$p=\omega \rho$, where $\omega$ is the barotropic constant factor, and found several approximate solutions which, however, turned out to be characterized by either negative values of the pressure or singularities at the origin. Although this type of solutions could have some mathematical interest, their physical interpretation is difficult due to the presence of singularities, negative pressures or negative energy densities. 

However, we investigated the structure of the field equations and found certain symmetries that allow us to generate new solutions from a given seed solution. This generating technique is based upon a transformation of the functions and coordinates entering the field equations. The result can be formulated as follows:

{\bf Theorem:} Let an exact interior solution of Einstein's equations (\ref{mu})-(\ref{gamt}) for the static axisymmetric line element (\ref{lel}) be given explicitly by means of the functions 
\be 
h_0 = h_0(r), \ \mu_0=\mu_0(r,\theta), \ \psi_0=\psi_0(r,\theta), \  \gamma_0=\gamma_0(r,\theta),\ \bar p_0 = \bar p_0(r,\theta), \ \bar\rho_0 = \bar\rho_0(r,\theta) \ .
\label{sol}
\ee 
Then, for any arbitrary real value of the constant parameter $\delta$ the new functions 
\be
h=h_0(r)  ,\  \mu = \mu_0(r,\tilde \theta) , \ \psi = \delta \psi_0(r,\tilde\theta) , 
\ \bar p = \delta \bar p_0(r,\tilde\theta) , \ \bar\rho = \delta \bar\rho_0(r,\tilde\theta), \ \tilde\theta = \frac{\theta}{\sqrt{\delta}} \ ,
\label{nsol}
\ee
\bea
\gamma(r,\tilde\theta)  = &&\delta^2 \gamma_0(r,\tilde\theta) + (\delta^2 -1 ) \int \frac{\nu_{\tilde\theta}}{h_0+\nu^2} dr\nonumber\\
&& + 8\pi \delta(1-\delta) \int \frac{ \frac{\mu_0}{\mu_{0,r}} \bar p_0}{h_0+\nu^2} dr + \kappa, \quad 
\nu =\frac{\mu_{0,\tilde\theta}}{\mu_{0,r}} \ ,\ \kappa= {\rm const.} 
\label{ngam}
\eea
represent a new solution of the field equations (\ref{mu})-(\ref{gamt}).

{\it Proof:} 
Since the set of functions (\ref{sol}) represents a solution, the field equations (\ref{mu}) and (\ref{psi})
\be
\frac{\mu_{0,rr}}{\mu_0} +  \frac{\mu_{0,\theta\theta}}{h_0\mu_0} +\frac{h_{0,r}\mu_{0,r}}{2h_0\mu_0} = \frac{16\pi}{h_0} \bar p_0 \ ,
\label{musol}
\ee
\be
\psi_{0,rr} + \frac{\psi_{0,\theta\theta}}{h_0} +\left(\frac{h_{0,r}}{2h_0}+\frac{\mu_{0,r}}{\mu_0}\right) \psi_{0,r} + \frac{\mu_{0,\theta} \psi_{0,\theta}}{h_0\mu_0} 
= \frac{4\pi}{h_0}(3 \bar p_0 + \bar\rho_0)\ ,
\label{psisol}
\ee
are identically satisfied. Notice that  the normalized quantities $\bar p_0$ and $\bar\rho_0$, as written in the above identities, are just explicitly known functions of their arguments so that we can treat them as independent functions. In other words, at the level of the  identities, which follow  from the field equations for the seed functions, we can treat  
$\bar p_0$, $\bar\rho_0$, $\psi_0$ and $\gamma_0$ as algebraically independent functions.  

If we now replace the functions $h=h_0/\delta$ and $\mu=\mu_0(r,\tilde\theta)$ in Eq.(\ref{mu}), and perform the coordinate reparametrization $\theta \rightarrow \tilde \theta = \theta/\sqrt{\delta}$, we obtain the equation 
\be
\frac{\mu_{0,rr}}{\mu_0} +  \frac{\mu_{0,\tilde\theta\tilde\theta}}{h_0\mu_0} +\frac{h_{0,r}\mu_{0,r}}{2h_0\mu_0} = \frac{16\pi}{h_0}\delta \bar p\ .
\label{munsol}
\ee
This equation coincides with the identity (\ref{musol}) only if $\tilde\theta \rightarrow \theta$ and $\delta \bar p \rightarrow \bar p_0$. So, in general, it is not identically satisfied. However, one can interpret it as the definition of a new pressure $\bar p_{new}(r,\tilde\theta) = \delta \bar p(r,\tilde\theta)$. To establish the connection with the seed solution we assume that $\bar p_{new}(r,\tilde\theta) = \delta \bar p_0(r,\tilde\theta)$. It then follows that the functions 
$h=h_0$ and $\mu=\mu_0(r,\tilde\theta)$ satisfy the field equation (\ref{mu}) for a new pressure $\delta \bar p_0(r,\tilde\theta)$. The next step is to show that these functions can be made to satisfy the remaining field equations. 

Consider the field equation (\ref{psi}) for $\psi(r,\tilde\theta)$ with $\mu = \mu_0(r,\tilde\theta)$, $h=h_0(r)$ and the new pressure $\bar p(r,\tilde\theta) =\delta \bar p_0(r,\tilde\theta)$. 
Then, we obtain
\be
\psi_{,rr} + \frac{\psi_{\tilde\theta\tilde\theta}}{h_0} +\left(\frac{h_{0,r}}{2h_0}+\frac{\mu_{0,r}}{\mu_0}\right) \psi_{,r} + \frac{\mu_{0,\tilde\theta} \psi_{\tilde\theta}}{h_0\mu_0} 
= \frac{4\pi}{h_0} (3 \delta \bar p_0 + \bar\rho)\ .
\label{npsi}
\ee
A simple inspection of this equation shows that it reduces to the identity (\ref{psisol}) for the choice $\psi=\delta \psi_0(r,\tilde\theta)$ and $\bar\rho = \delta \bar\rho_0(r,\tilde\theta)$. This proves that the functions (\ref{nsol}) determine a new solution of the field equations (\ref{mu}) and (\ref{psi}).

We now calculate the explicit form of the function $\gamma$. First of all, we notice that since $\gamma_0(r,\theta)$ is a solution so is also $\gamma_0(r,\theta)$ when we change $\theta$ by $\tilde\theta$ everywhere in the equations. Then, the corresponding field equations become identities, i. e.,
\be
\gamma_{0,r} = \frac{1}{H(r,\tilde\theta )}\bigg\{\mu_0\left[\mu_{0,r}\left(h_0\psi_{0,r}^2-\psi_{0,\tilde\theta }^2\right) +2\mu_{0,\tilde\theta }\psi_{0,\tilde\theta }\psi_{0,r} 
+  8\pi\mu_{0,r}\bar p_0\right]
+\mu_{0,\tilde\theta }\mu_{0,r\tilde\theta }-\mu_{0,r}\mu_{0,\tilde\theta \tilde\theta }  \bigg\}\ ,
\label{gamr0}
\ee
\be
\gamma_{0,\tilde\theta } = \frac{1}{H(r,\tilde\theta )}\bigg\{ \mu_0\left[\mu_{0,\tilde\theta }\left(\psi_{0,\tilde\theta }^2-h_0\psi_{0,r}^2\right) 
+2 h_0 \mu_{0,r}\psi_{0,\tilde\theta }\psi_{0,r}  - 8\pi\mu_{0,\tilde\theta }\bar p_0 \right] 
+ h_0\mu_{0,r}\mu_{0,r\tilde\theta } +\mu_{0,\tilde\theta }\mu_{0,\tilde\theta \tilde\theta }\bigg\}\ ,
\label{gamt0}
\ee
with
\be
H(r,\tilde\theta ) = h_0\mu_{0,r}^2+\mu_{0,\tilde\theta }^2\ .
\ee
We now introduce the new solution (\ref{nsol}) into the field equation (\ref{gamr}), and obtain 
\bea
\gamma_{,r} = \frac{1}{H(r,\tilde\theta) }  \bigg\{ & \delta^2\mu_0\left[\mu_{0,r}\left(h_0\psi_{0,r}^2-\psi_{0,\tilde\theta}^2\right) 
+2\mu_{0,\tilde\theta}\psi_{0,\tilde\theta}\psi_{0,r} 
\right] \nonumber\\
& +  8\pi\delta \mu_0 \mu_{0,r}\bar p_0 +  \mu_{0,\tilde\theta}\mu_{0,r\tilde\theta}-\mu_{0,r}\mu_{0,\tilde\theta\tilde\theta}  \bigg\}\ .
\eea
Using Eq.(\ref{gamr0}) to replace the term in squared brackets of  the last equation, after some algebraic rearrangements, we get
\be
\gamma_{,r} = \delta^2 \gamma_{0,r} + \frac{\delta^2-1}{H(r,\tilde\theta)} \left(\frac{\mu_{0,\tilde\theta}}{\mu_{0,r}}\right)_{\tilde\theta} \mu_{0,r}^2 
+ 8\pi \delta(1-\delta)\frac{\mu_0\mu_{0,r}\bar p_0}{H(r,\tilde\theta)}\ .
\label{gamrn}
\ee 

Consider now the second equation (\ref{gamt}) for the function $\gamma$. Introducing the values of the new solution (\ref{nsol}) into Eq.(\ref{gamt}), we obtain
\bea
\gamma_{,\tilde\theta} = \frac{1}{H(r,\tilde\theta)} \bigg\{ & \delta^{2}\mu_0\left[\mu_{0,\tilde\theta}\left(\psi_{0,\tilde\theta}^2-h_0\psi_{0,r}^2\right) 
+2 h_0 \mu_{0,r}\psi_{0,\tilde\theta}\psi_{0,r} \right] \nonumber\\ 
&   - 8\pi\delta \mu_0\mu_{0,\tilde\theta}\bar p_0 + \left( h_0\mu_{0,r}\mu_{0,r\tilde\theta} +\mu_{0,\tilde\theta}\mu_{0,\tilde\theta\tilde\theta}\right) \bigg\}\ .
\eea
Furthermore, we  use the identity (\ref{gamt0}) to replace the expression contained in the squared brackets. It is then straightforward to get
\be
\gamma_{,\tilde\theta} = \delta^2 \gamma_{0,\tilde\theta} + \frac{1}{2}(1-\delta^2) \frac{\partial}{\partial\tilde\theta} \ln H(r,\tilde\theta)
+8\pi \delta(\delta-1) \frac{\mu_0\mu_{0,\tilde\theta} \bar p_0 }{H(r,\tilde\theta)}
\ .
\ee
The above equation can immediately be integrated, and the solution is determined up to an arbitrary function of $r$ 
\be
\gamma(r,\tilde\theta) = \delta^2\gamma_0 + \frac{1}{2} (1-\delta^2) \ln H
+8\pi\delta(\delta-1) \int\frac{\mu_0\mu_{0,\tilde\theta} \bar p_0}{H(r,\tilde\theta)} d\tilde\theta 
 + F(r) 
\label{gamn0}
\ee
which can be fixed by considering the second equation (\ref{gamrn}) for the function $\gamma(r,\tilde\theta)$. We obtain 
\bea  
F(r) = & & (1-\delta^2) \left[ \int \frac{ \mu_{0,\tilde\theta}\mu_{0,r\tilde\theta}-\mu_{0,r}\mu_{0,\tilde\theta\tilde\theta} }{h_0\mu_{0,r}^2+\mu_{0,\theta}^2} dr
- \frac{1}{2} \ln H(r,\tilde\theta) \right] \nonumber\\ 
& + & 8\pi\delta(1-\delta) \left[ \int \frac{\mu_0\mu_{0,\tilde\theta} \bar p_0}{H(r,\tilde\theta)} d\tilde \theta
+ \int \frac{\mu_0\mu_{0,r} \bar p_ 0}{H(r,\tilde\theta)} dr \right]
+ \kappa \ ,
\eea
where $\kappa$ is an integration constant. 
Inserting this expression into Eq.(\ref{gamn0}), after some algebraic rearrangements we  finally obtain
\be
\gamma(r,\tilde\theta) = \delta^2 \gamma_0(r,\tilde\theta) + (\delta^2-1) \int \frac{\nu_{\tilde\theta}}{h_0 + \nu^2} dr 
+ 8\pi\delta(1-\delta) \int \frac{\mu_0 \mu_{0,r} \bar p_0 }{H(r,\tilde\theta)} dr +
\kappa \ ,
\ee
which is equivalent to Eq.(\ref{ngam}). 
This ends the proof of the theorem.

\section{Examples}
\label{sec:ex}

The theorem proved in the previous section states that given a static axisymmetric solution it is possible to generate a new solution which contains an additional new parameter $\delta$.  In this section, we will apply the transformation to show that it can be used to generate new solutions that are physically different from the seed solution.

\subsection{The vacuum q-metric}
\label{sec:qmet}

Let us consider the vacuum limiting case of the above transformation. The particular choice 
\be
p=0 ,\  \rho=0 ,\ \mu=r  , \ h=1, \ \theta = z 
\ee
in Eq.(\ref{lel}) leads to the static axisymmetric Lewis-Papapetrou line element for vacuum spacetimes \cite{solutions}
\be
ds^2 = e^{2\psi} dt^2 - e^{-2\psi}\left[ e^{2\gamma} (d r^2 + dz^2 ) + r^2 d\varphi^2\right]
\label{vaclel}
\ee
in cylindrical coordinates $(t,r,z,\varphi)$ for which the field equations (\ref{mu})-(\ref{gamt}) reduce to 
\be
\psi_{,rr} + \frac{1}{r} \psi_{,r} + \psi_{,zz} = 0 \ ,\ \gamma_{,r} = r(\psi_{,r}^2 - \psi_z^2) , \ \gamma_z = 2 r \psi_{,r}\psi_z\ ,
\label{vacfeq}
\ee
for the functions $\psi$ and $\gamma$ only. Then, according to the Theorem proved in the previous section, if $\psi_0 = \psi_0(r,z)$ and $\gamma_0 = \gamma_0(r,z)$ represent a particular solution of Eqs.(\ref{vacfeq}), a new solution $\psi$, $\gamma$ is given by
\be
\psi(r,\tilde z) = \delta \psi_0(r,\tilde z) , \ \gamma(r,\tilde z) = \delta^2 \gamma_0(r,\tilde z) \ .
\label{zvtra}
\ee
Notice that for this new solution the line element corresponds to (\ref{vaclel}) with $z\rightarrow \tilde z$. Since the new functions depend also on $\tilde z$, one can use the coordinate $z$ instead, without loss of generality.
The transformation (\ref{zvtra}) was originally proposed by Zipoy \cite{zip66}  and Voorhees \cite{voor70} and has been investigated in several works (see, for instance, \cite{par85,zv1,zv2,zv3,zv4,zv5,mala04,quev11} and references therein). For this reason, the transformation defined in Eqs.(\ref{nsol}) and (\ref{ngam}) can be considered as a generalization of the Zipoy-Voorhees transformation that includes the case of perfect-fluid spacetimes. 

A particular example of the Zipoy-Voorhees transformation arises if we take the Schwarzschild metric as seed solution as follows. First, we introduce spherical coordinates 
$(t, \tilde r, \vartheta, \varphi)$ into the line element (\ref{vaclel}) by means of the transformations 
\be
\tilde r = m + \frac{1}{2}(r_++r_-)\ , \quad \cos\vartheta = \frac{1}{2m}(r_+-r_-)\ , \quad r_\pm = \sqrt{r^2 + (z\pm m)^2} \ ,
\ee
and obtain 
\bea
ds^2 = && e^{2\psi} dt^2 - e^{-2\psi} \bigg[ e^{2\gamma}\left( 1 - \frac{2m}{\tilde r} + \frac{m^2\sin^2\vartheta}{\tilde r^2} \right) 
\left(\frac{d\tilde r ^2}{1-\frac{2m}{\tilde r}} + \tilde r ^2 d\vartheta^2\right) \nonumber\\
& + & \tilde r ^2\left(1-\frac{2m}{\tilde r} \right) \sin^2\vartheta d\varphi^2 \bigg]\ .
\eea
Then, the Schwarzschild metric corresponds to the particular solution
\be 
\psi_0 = \frac{1}{2}\ln \frac{\tilde r - 2 m}{\tilde r} \ , \quad 
\gamma_0 = \frac{1}{2}\ln \frac{\tilde r^2 - 2 m \tilde r}{\tilde r^2 - 2m\tilde r + m^2 \sin^2\vartheta}
\ .
\ee
Applying the transformation (\ref{zvtra}) with $\delta = 1 + q$, i.e.,
\be 
\psi =  \frac{1}{2}(1+q)\ln \frac{\tilde r - 2 m}{\tilde r} \ , \quad 
\gamma = \frac{1}{2} (1+q)^2 \ln \frac{\tilde r^2 - 2 m \tilde r}{\tilde r^2 - 2m\tilde r + m^2 \sin^2\vartheta}
\ ,
\ee
the resulting line element can be expressed as 
\bea
ds^2 = & & \left(1-\frac{2m}{\tilde r}\right)^{1+q} dt^2 \nonumber  \\
& -& \left(1-\frac{2m}{\tilde r}\right)^{-q}\left[ \left(1+\frac{m^2\sin^2\vartheta}{\tilde r^2-2m\tilde r}\right)^{-q(2+q)} \left(\frac{d\tilde r^2}
{1-\frac{2m}{\tilde r}}+ \tilde r^2d\vartheta^2\right) + \tilde r^2 \sin^2\vartheta d\varphi^2\right] ,
\label{zv}
\eea
which, as expected, reduces to the Schwarzschild metric in the limiting case $q\rightarrow 0$.

A detailed analysis 
 of this metric shows that $m$ and   $q$ are constant parameters that determine the total mass and the quadrupole moment of the gravitational source  \cite{quev11}.
The metric (\ref{zv}) has been interpreted as the simplest generalization of the Schwarzschild metric with a quadrupole. In the literature, this metric is known as 
the $\delta-$metric, referring  to the parameter $\delta$,  or as the $\gamma-$metric for notational reasons. We 
use the term $q-$metric to emphasize the physical interpretation of the new solution in terms of the quadrupole moment. 

Whereas the  seed metric is the spherically symmetric Schwarzschild solution, which describes the gravitational field of a black hole, the generated $q-$metric is axially symmetric and describes the exterior field of a naked singularity \cite{quev11}. This shows that the transformation presented in the previous section generates non-trivial solutions even in the case of vacuum solutions.

\subsection{Generalization of the interior Schwarzschild solution}
\label{sec:schwgen}

One of the most important interior solutions of Einstein's equations is the  spherically symmetric Schwarzschild solution 
which describes the interior field of a perfect-fluid sphere of radius $R$ and total mass $m$. The corresponding line element can be written as
\be
ds^2 = \left[\frac{3}{2}f(R) - \frac{1}{2}f(r)\right]^2 dt^2 - \frac{dr^2}{f^2(r)} - r^2(d\theta^2 + \sin^2\theta d\varphi^2)\ ,
\ee
with 
\be
f(r)= \sqrt{1-\frac{2mr^2}{R^3}}\ .
\ee
The parameters of the perfect fluid are the constant density $\rho_0$ and the pressure $p_0$ which is a function of the radial coordinate $r$ only 
\be
p_0 = \rho_0 \frac{f(r) - f(R)}{3f(R)-f(r)}
\ee

We will use the above metric as the seed solution (\ref{sol}) for the general transformation (\ref{nsol}). Then, a straightforward comparison with the general line element (\ref{lel}) yields
\be
e^{\psi_0} = \frac{3}{2}f(R) - \frac{1}{2}f(r)\ , \ h_0 = r^2 f^2(r) \ , \ \mu_0 = r \sin\theta e^{\psi_0}\ , \ e^{\gamma_0} = r e^{\psi_0} \ .
\label{schint1}
\ee
According to the Theorem presented in Sec.\ref{sec:tra}, the new solution can be obtained from Eq.(\ref{schint1}) by multiplying the corresponding metric functions with the new parameter $\delta$. Then, the new line element can be represented as
\be
ds^2 = e^{2\delta \psi_0}   dt^2 -    e^{- 2\delta \psi_0}    
         \left[e^{2\gamma} \left(\frac{dr^2}{r^2f^2(r)} + d\tilde \theta ^2 \right) + r^2 e^{2\psi_0}\sin^2 \tilde \theta  d\varphi^2\right] \ ,
\ee
where the new function $\gamma$ is given by
\be
\gamma  =   \delta^2 \gamma_0  + \int \frac{ (1-\delta^2) +  8\pi\delta (1-\delta)\sin^2\tilde\theta r^2 p_ 0 }
{  r f^2(r) (1+r\psi_{0\,r})\sin^2\tilde \theta + \frac{r}{ 1+r\psi_{0\,r}}\cos^2\tilde\theta } dr
 + \kappa\ ,
\label{ngs}
\ee
where 
\be
\psi_{0\,r} = \frac{2mr}{R^3 f(r)[3f(R)-f(r)]}\ .
\ee

Moreover, the parameters of the perfect fluid source are
\be
\rho = \delta \rho_0 e^{2\gamma_0-2\gamma + 2(\delta-1) \psi_0}\ \ , p = \delta p_0 e^{2\gamma_0-2\gamma + 2(\delta-1) \psi_0}\ .
\ee
Notice that the new function $\gamma$ as given in Eq.(\ref{ngs}) depends explicitly  on the coordinate $\tilde\theta$, in contrast to the seed metric function  $\gamma_0$ which depends on the coordinate $r$ only. This proves that the new solution is not spherically symmetric, indicating that it is physically  different from the interior Schwarzschild solution. This conclusion is corroborated also by the fact that the density and pressure of the new solution are functions of the angular coordinate too.

\section{Conclusions}
\label{sec:con}

In this work, we presented a new method for generating  perfect-fluid solutions of the Einstein equations, 
starting from a given seed solution. The method is based upon the introduction of a new parameter at the level of the
metric functions of the seed solution in such a way that the generated new solution is characterized by physical properties which are different from those of the seed solutions. This means that the solution generating technique proposed in this work can be used to generate non-trivial new solutions. 

To show the validity of the method we propose a line element which is especially adapted to handle the problem. In fact, the 
perfect-fluid field equations turn out to be split into a set of two partial differential equations that once solved can be used to integrate by quadratures the remaining equations. We also introduce in the line element an auxiliary function that is not fixed by the field equations and, therefore, can be used to simplify the analysis of the main field equations. 

The method has a well-defined vacuum limit. In fact, for a particular choice of the metric functions and for vanishing density and pressure, we obtain the Lewis-Papapetrou line element for static axisymmetric vacuum fields. In this limit, the generating technique proposed here coincides with the Zipov-Voorhees transformation which can be used to generate vacuum solutions from vacuum solutions. As a particular example we derive from the Schwarzschild metric a solution with quadrupole moment, which we  call 
the $q-$metric, and can be interpreted as describing the simplest generalization of the Schwarzschild spacetime with a quadrupole parameter. 

In the case of perfect-fluid solutions, we use the interior spherically symmetric Schwarzschild solution with constant density to generate a new interior solution which turns out to be axially symmetric. 

The generation of further new solutions does not represent any particular difficulty for the method developed in this work. The important problem of matching the so generated interior solutions with the corresponding exterior solutions is beyond the scope of the present work, and will be the subject of future investigations.

\section*{Acknowledgements}

This work was  supported by DGAPA-UNAM, Grant No. 113514, Conacyt-Mexico, Grant No. 166391, and 
MES-Kazakhstan, Grant No. 3098/GF4.   


\end{document}